\pdfoutput=1
\documentclass{JINST}
\let\ifpdf\relax
\usepackage{upgreek}
\title{Design of the AGIPD Sensor for the European XFEL}

\author{ J.~Schwandt$^{a,}$\thanks{Corresponding author.}~, E.~Fretwurst$^a$, R.~Klanner$^a$ and J.~Zhang$^a$\\
\llap{$^a$}Institute for Experimental Physics, University of Hamburg\\
  Luruper Chaussee 149, D-22761 Hamburg, Germany\\
  E-mail: \email{joern.schwandt@desy.de}}

\abstract{
 For experiments at the European X-Ray Free-Electron Laser (XFEL) the Adaptive Gain Integrating Pixel Detector (AGIPD) is under development.
 The particular requirements include a high dynamic range of 0,~1 to more than $10^{4}$ photons of 12.4~keV  per pixel within an XFEL pulse of less than 100~fs duration, and a radiation tolerance for doses up to 1~GGy for 3~years of operation.
 AGIPD is a hybrid-pixel detector system with a total of 1024 $\times$ 1024 pixels of size 200~$\upmu$m $\times$ 200 $\upmu$m.
 AGIPD will use 16~$p^{+}n$-silicon sensors of 500~$\upmu$m thickness.
 The nominal operating voltage is 500~V, however, for special applications operation close to 1000~V should be possible.
 Experimental data on the X-ray-dose dependence of the oxide-charge density and of the surface-current density have been implemented in the SYNOPSYS TCAD simulation program in order to optimize the design of the pixel and guard-ring layout.
 The methodology of the sensor design, the optimization of the most relevant parameters, and the optimized layout are described.
 Finally, the simulated performance, in particular the breakdown voltage, dark current and inter-pixel capacitance as function of the X-ray dose are presented.
}


\begin{document}

\section{Introduction}
 For experiments at the European X-Ray Free-Electron Laser (XFEL) the Adaptive Gain Integrating Pixel Detector (AGIPD) is under development \cite{Altarelli:2007,Henrich:2011}.
 It is a hybrid-pixel detector system with 1024 $\times$ 1024 $p^{+}$-pixels of dimensions 200 $\upmu$m $\times$ 200 $\upmu$m, built of 16  $p^+n$-silicon sensors, each with a sensitive area of 10.52~cm $\times$ 2.56~cm and a thickness of 500 $\upmu$m.
 The particular requirements include a dynamic range of 0,~1 to more than 10$^4$ photons of 12.4~keV per pixel for a pulse duration of less than 100~fs, negligible pile-up at the XFEL repetition rate of 4.5$~$MHz, and operation for X-ray doses up to 1~GGy in 3 years \cite{Graafsma:2009}.

 The paper describes the optimization of the AGIPD sensor, its final design and expected performance.
 After a short summary of results on  X-ray radiation damage, the methodology of the optimization of the guard-ring structure and of the pixel layout using the SYNOPSYS TCAD~\cite{Synopsys} simulation program is presented.
 Finally, the optimized technological parameters and sensor layout, as well as the expected sensor performance, in particular the breakdown voltage, dark current and inter-pixel capacitance as function of X-ray dose are discussed.
 Based on the results of the simulations it is concluded that the AGIPD specifications can be met.

\section{X-ray radiation damage of segmented silicon sensors and TCAD simulation}

 For 12.4~keV photons the maximum energy transfer to silicon atoms is well below the threshold energy of 21~eV for bulk damage \cite{Akkerman:2001}.
 Therefore, no bulk damage is expected, however surface damage will occur. The X-rays produce electron-hole pairs in the SiO$_2$. Whereas most of the electrons will leave the oxide, the holes, which have a much lower mobility than the electrons, will produce positive charges in the oxide or Si-SiO$_2$-interface states~\cite{Oldham:1999}.


 In the case of a $p^+n$ sensor the radiation-induced positive oxide charge will produce a sharp bending of the $p^+$-depletion boundary near the Si-SiO$_2$ interface, resulting in a high electric field and a reduced breakdown voltage.
 In addition, an electron-accumulation layer forms, which prevents the depletion of the silicon below the SiO$_2$ and which causes an increase of the inter-pixel capacitance and charge losses.

 The charge states of the Si-SiO$_2$ interface traps with energies in the Si-band gap depend on the trap type, their energies and the position of the Fermi level.
 They are responsible for the surface-generation current from regions where the Si-SiO$_2$ interface is depleted and thus exposed to an electric field.

\begin{figure}[htpb]
  \centering
  \includegraphics[width=0.45\linewidth]{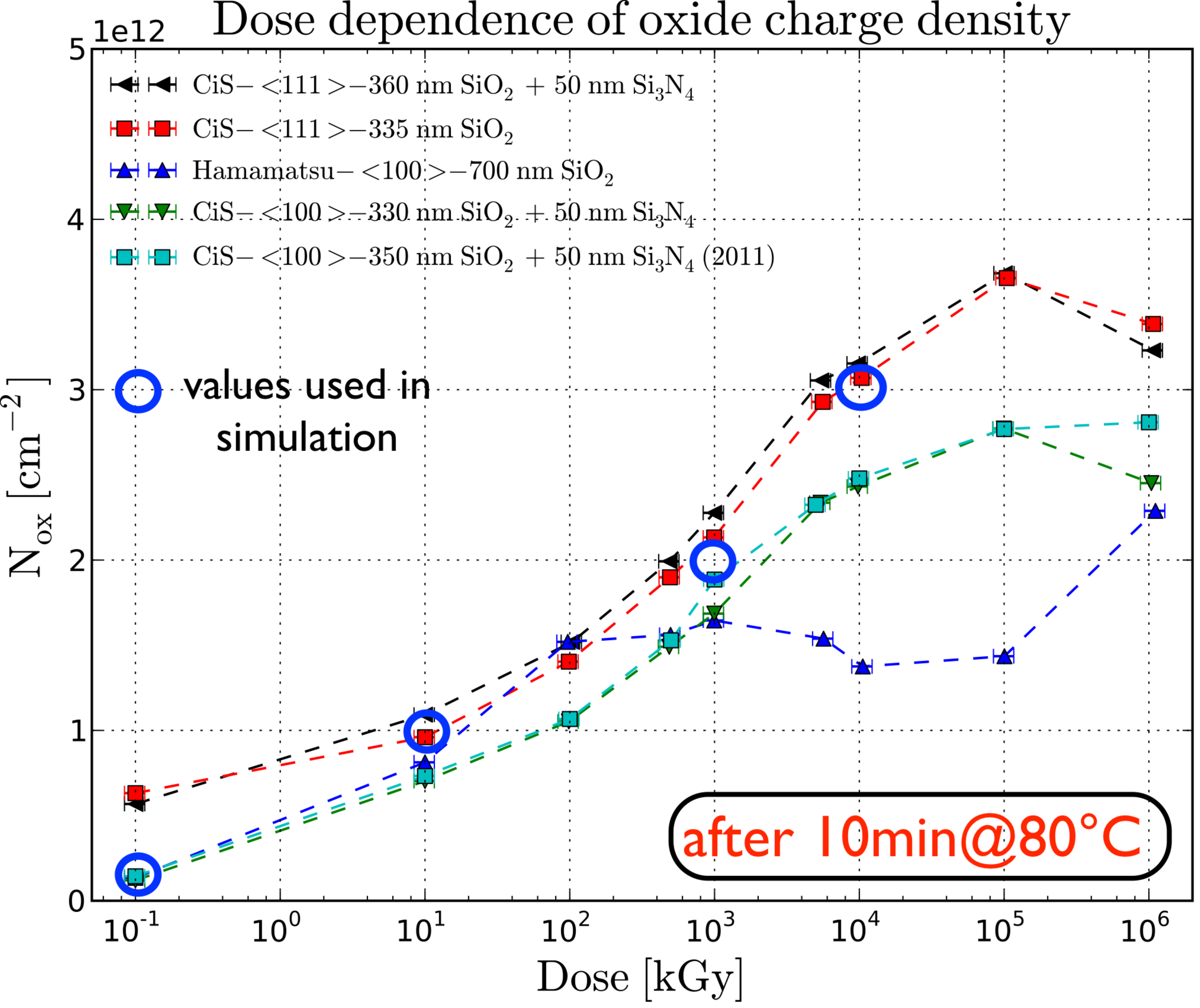}
  \hspace{0.5cm}
  \includegraphics[width=0.45\linewidth]{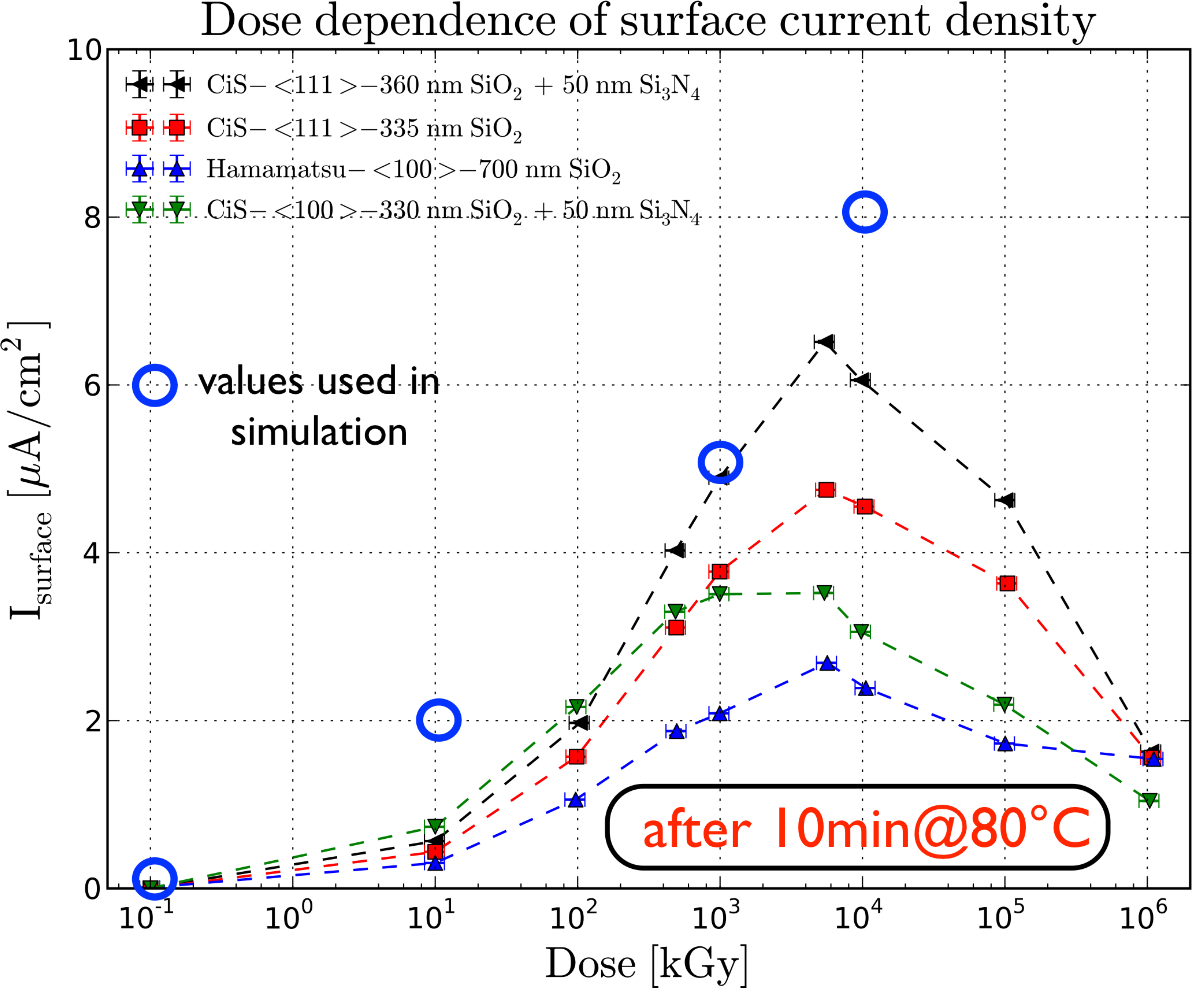}
  \caption{Dependence of the surface density of oxide charges (left) and
  surface-current density (right) on the X-ray irradiation dose for test structures
  (MOS capacitors, gate-controlled diodes) from different vendors after annealing
  at 80~$^\circ$C for 10 minutes.}
  \label{fig:surfdamdata}
\end{figure}

 Measurements on test structures (MOS capacitors and gate-controlled diodes) have been used to determine the dose dependence of the density of oxide charges $N_{ox}$, and the surface-current density $J_{surf}$.
 The results are shown in figure \ref{fig:surfdamdata} \cite{Zhang:2011a,Zhang:2012}.
 In order to obtain reproducible results after irradiation, the test structures had to be annealed for 10 minutes at 80$^\circ$C~\cite{Zhang:2011a}.
 It is found that $N_{ox}$ increases with dose and saturates for doses above $\approx$~10~MGy.
 The values for MOS capacitors from different vendors and for different crystal orientations show a spread of about a factor~2.
 In addition, annealing is observed already at room temperature \cite{Zhang:2012}.
 $J_{surf}$ also increases up to $\approx$ 10~MGy and then drops to lower values. The reason for this drop is not understood and still under study.

 The values which were used in the TCAD simulations are indicated in the figure as circles.
 In the simulations it was assumed that $N_{ox}$ is uniformly distributed along the Si-SiO$_2$ interface.
 For the generation of the surface current the surface-recombination velocity $s_{0} = J_{surf}/ (0.5\cdot q_{0}\cdot n_{i})$ with the elementary charge $q_{0}$ and the intrinsic charge-carrier density $n_{i}$, was used.
 Details on the models for the device simulation and for the scaling of the 2D results to 3D for the pixels, can be found in \cite{Schwandt:2012}.
 The breakdown voltage was determined by the criterion
  \begin{displaymath}
    \frac{dI}{dV} / \frac{I}{V} = 10\; ,
  \end{displaymath}
 with the gradient of the current-voltage curve $dI/dV$, and the ratio of the current to voltage $I/V$.

\section{Sensor optimization}
 Based on detector and science simulations \cite{Potdevin:2009}, the AGIPD collaboration has specified the sensor parameters to be achieved for doses from 0 to 1~GGy listed in table~\ref{tab:specs}.

 \begin{table}[htpb]
	\begin{center}
		\small
		\begin{tabular}{|c|c|c|}\hline
           Parameter & Value \\ \hline \hline
           sensitive area & 10.52 $\times$ 2.56 cm$^2$ \\ \hline
           mechanical thickness & 500 $\pm$ 20 $\upmu$m \\ \hline
           distance pixel edges to cut edges & 1200 $\upmu$m \\ \hline
           resistivity of $n$ doping & 3 - 8 k$\Omega\cdot$cm \\ \hline
           pixel dimensions & 200 $\times$ 200 $\upmu$m$^2$ \\ \hline
           operating voltage & 500 V \\ \hline
           breakdown voltage & > 900 V \\ \hline
           coupling type & DC \\ \hline
           inter-pixel capacitance@500V & < 500 fF \\ \hline
           total sensor dark current@500V & < 50 $\upmu$A \\ \hline
		\end{tabular}
		\caption{Specifications for the AGIPD sensor for X-ray doses between 0 and 1~GGy}
		\label{tab:specs}
	\end{center}
\end{table}

\begin{figure}[htpb]
  \centering
  \includegraphics[width=0.45\linewidth]{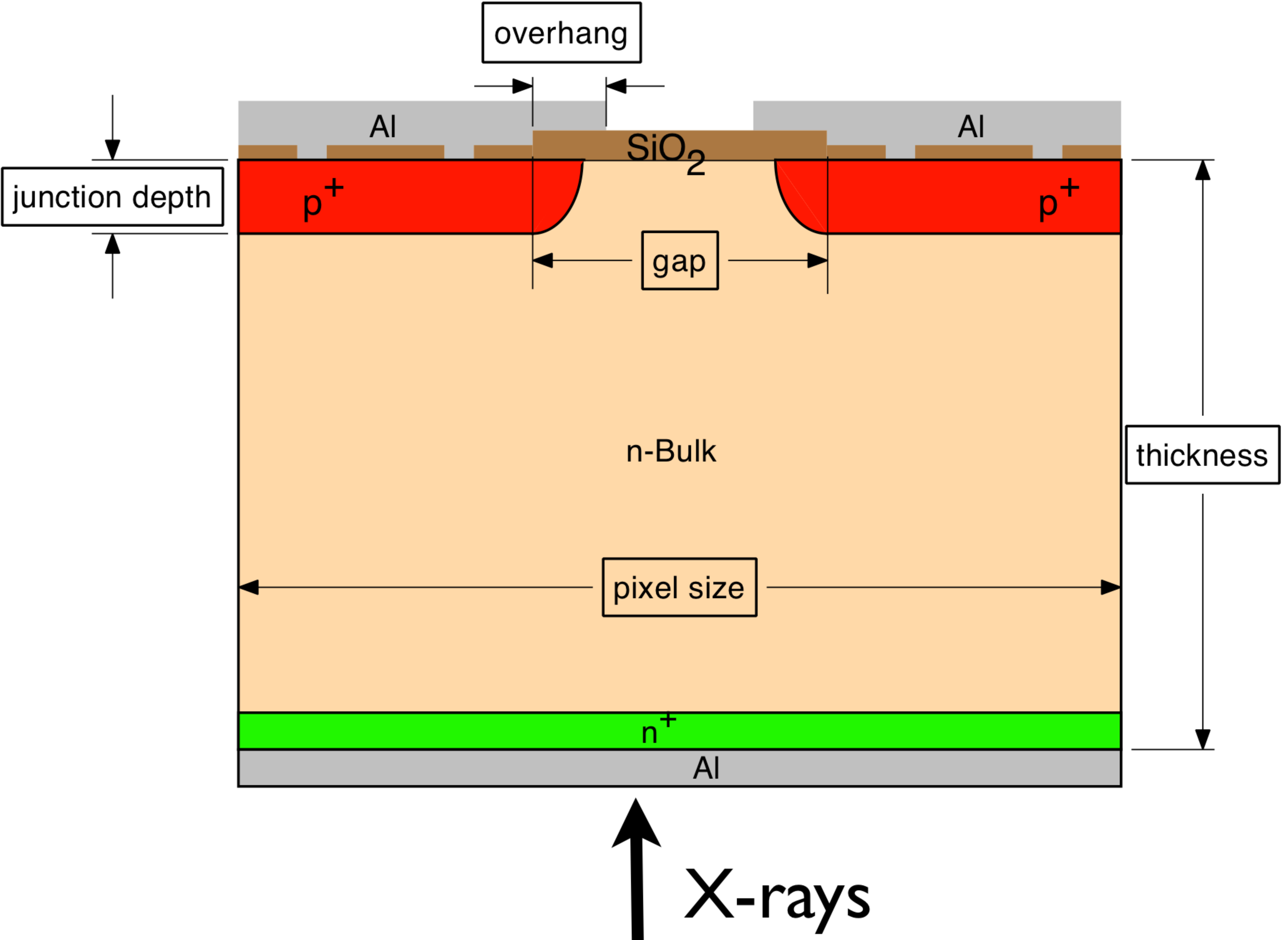}
  \hspace{0.5cm}
  \includegraphics[width=0.45\linewidth,height=13em]{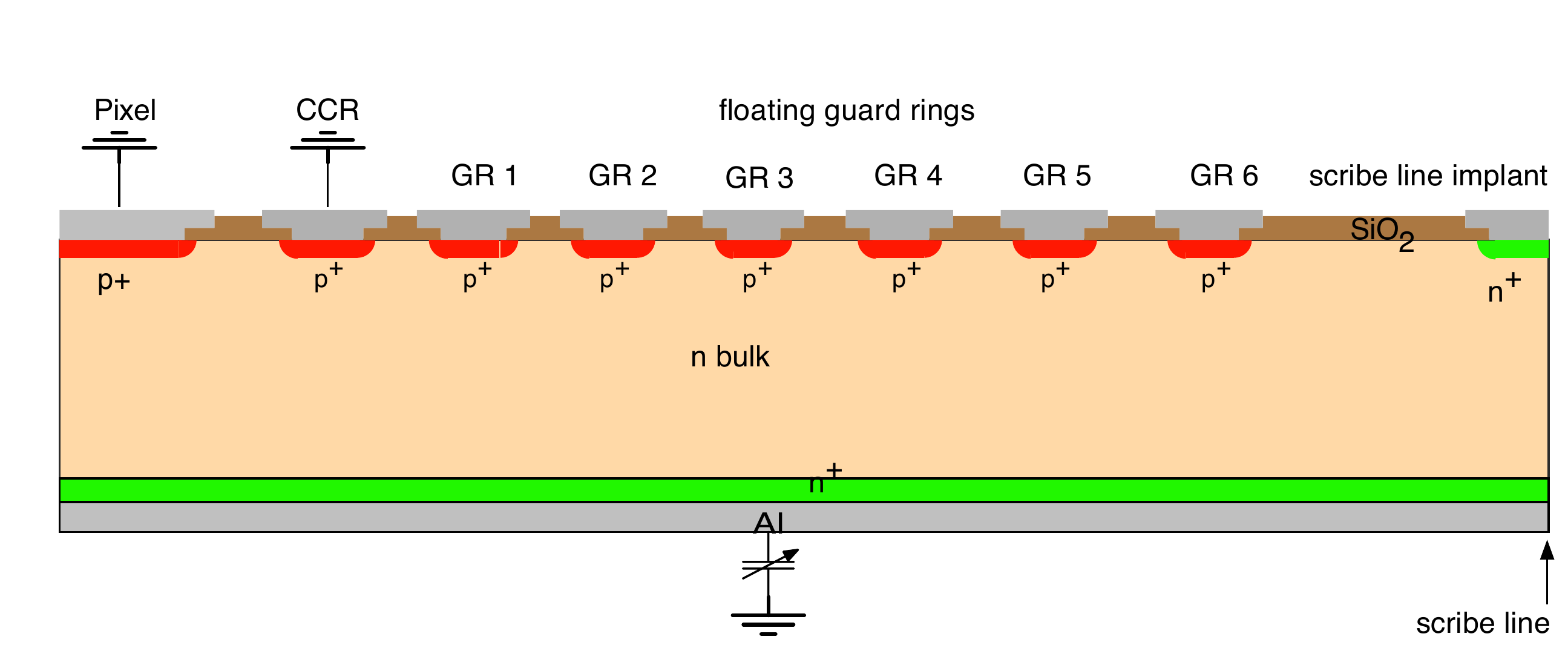}
  \caption{Left: Sketch of the sensor region simulated for the pixel optimization.
           Right: The region used for the guard ring simulation consists of half a pixel, the current collection ring (CCR), floating guard rings (GR - only 6 are shown) and the $n^+$ scribe-line implant. Backplane and  scribe-line implant are assumed to be on the same potential.}
  \label{fig:sketch}
\end{figure}

 The parameters which have been optimized in the design of the sensor, some which are shown in figure~\ref{fig:sketch}, are for the pixels the gap, aluminum (Al) overhang and radius of $p^+n$ junction and Al layer at the corners, for the guard-ring structure the number of rings, implantation widths, spacings and Al overhangs, and for the process parameters the depth of the junctions and the oxide thickness.

 The performance parameters which have been optimized for doses between 0 and 1~GGy are  the breakdown voltage, the leakage current and inter-pixel capacitance for the pixels, and the  breakdown voltage and the voltage drop between the individual guard rings for the guard-ring structure.
 Moreover, it has been ensured that the depletion boundary does not touch the scribe line.\\

 The strategy applied for the optimization can be summarized as follows:
\begin{itemize}
\item Guard ring (GR) optimization (2D simulations in (x,y) for the straight edges and in (r,z) for the corners):
  \begin{enumerate}
   \item Optimize the breakdown voltage $V_{bd}$ without guard ring (0~GR - only CCR)
    for different values of $N_{ox}$
   ($5\times 10^{10}$, $10^{11}$, $10^{12}$, $2\times10^{12}$, and $3\times 10^{12}$ cm$^{-2}$) as
    function of oxide thickness, junction depth and Al overhang.
   \item Estimate the number of floating GRs required for $V_{bd} = 1000$~V.
   \item Vary the spacing between GRs, junction widths and Al overhangs, to achieve the maximum $V_{bd}$ for a minimum width of the GR structure. This was achieved when the maximum electric fields in the gaps between the guard rings are approximately the same.
  \end{enumerate}
\item Pixel optimization (2D simulations in (x,y)):
  \begin{enumerate}
   \item Optimize the oxide thickness, Al overhang, gap and junction depth with respect to breakdown voltage, dark current and inter-pixel capacitance.
   \item Extrapolate the dark current and inter-pixel capacitances to 3D values.
   \item Check the breakdown voltage and dark current as function of voltage by a 3D simulation (only 1/4 pixel simulated).
  \end{enumerate}
\end{itemize}

\section{Results of the sensor optimization}

\subsection{Guard-ring structure}

\begin{figure}[htpb]
  \centering
  \includegraphics[width=0.8\linewidth]{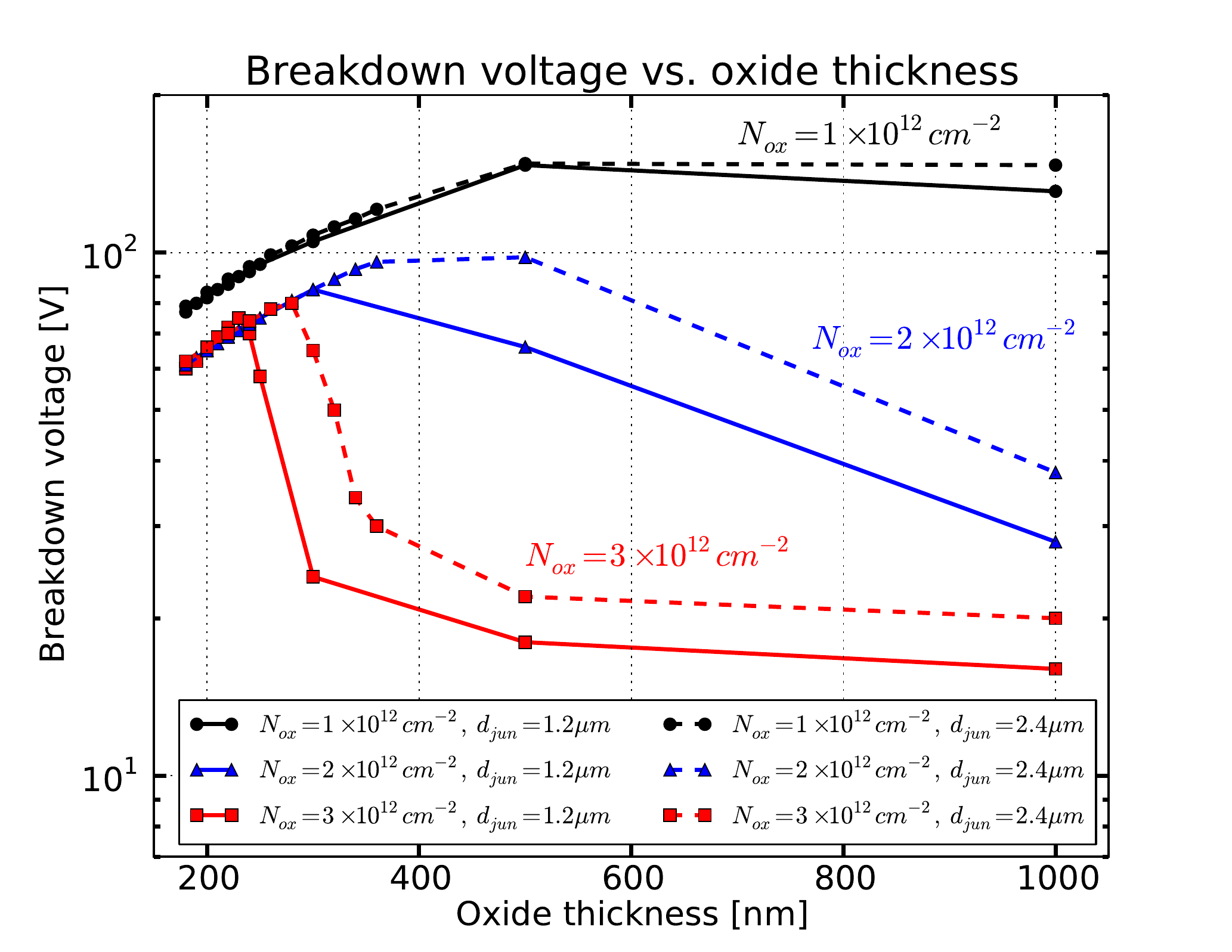}
  \caption{Breakdown voltage for 0 GR as function of the oxide thickness for different values of $N_{ox}$. The results are for a resistivity of 5~k$\Omega\cdot$cm, an Al overhang of 5~$\upmu$m and junction depths $d_{jun}$ of 1.2~$\upmu$m (solid) and 2.4~$\upmu$m (dashed). The breakdown voltages for $N_{ox}$ values below $10^{12}$ cm$^{-2}$ are above 180~V.}
  \label{fig:bdvsox}
\end{figure}

  The breakdown voltage $V_{bd}$ without guard ring (0~GR) as function of oxide thickness $d_{ox}$, for different oxide-charge densities $N_{ox}$, two junction depths $d_{jun} = 1.2$ and 2.4~$\upmu$m and an Al overhang of 5~$\upmu$m, is shown in figure~\ref{fig:bdvsox}.
  The breakdown voltages for $N_{ox}$ values below $10^{12}$~cm$^{-2}$ are above 180~V and are not shown.
  For $N_{ox} = 10^{12}$~cm$^{-2}$ the values of $V_{bd}$ increase from 80~V to 150~V when $d_{ox}$ is increased.
  For higher $N_{ox}$~values, $V_{bd}$ reaches a maximum and then decreases.
  For $N_{ox} = 3\times 10^{12}$~cm$^{-2}$ the value of $V_{bd}$ drops from 70~V at $d_{ox} = 250$~nm, to 20~V at $d_{ox} = 500$~nm.
  This sudden decrease of $V_{bd}$ is related to the electron-accumulation layer below the Al overhang: 
  If an accumulation layer is present, there is a single high-field region at the edge of the $p^+n$~junction.
  Without accumulation layer the voltage drop occurs over the entire region below the Al layer, with one high field region at the edge of the junction and a second one below the end of the Al overhang.
  In addition, a deeper $p^+n$~junction results in a lower field at the edge of the junction, thus further increasing the breakdown voltage.

  From the curves of $N_{ox} = 3\times 10^{12}$ cm$^{-2}$ shown in figure~\ref{fig:bdvsox}, we conclude that the optimum value of the oxide thickness is 230~nm for a junction depth of 1.2~$\upmu$m, and 270~nm for 2.4~$\upmu$m.
  The corresponding breakdown voltages are between 70 and 80~V\@.
  This leads to the estimate that 15 floating GRs and a CCR are required to achieve a sensor-breakdown voltage of $\approx 1000$~V\@.

 After choosing a junction depth of 2.4~$\upmu$m, an oxide thickness of 250~nm, and performing step 3 of the GR-optimization procedure, the following parameters for the optimized guard-ring structure have been determined:
\begin{itemize}
 \item Gap between pixels and CCR: 20 $\upmu$m.
 \item Width of the $p^+n$~junction of the CCR: 90 $\upmu$m.
 \item Al overhang of CCR: 5 $\upmu$m.
 \item Gap between CCR and 1st GR: 12~$\upmu$m.
 \item Widths of the $p^+n$ junctions of the GRs: 25~$\upmu$m.
 \item Al overhangs towards pixel for GRs 1-15: 2, 3, 4,..., 16 $\upmu$m.
 \item Al overhangs away from pixel for GRs 1-15: 5 $\upmu$m.
 \item Gaps between GRs 1-2,2-3,...,14-15: 12, 13.5, 15, ... 33 $\upmu$m.
\end{itemize}
 A printout of the GDS file of the optimized guard-ring design is shown in figure~\ref{fig:gds} left.
 The simulated breakdown voltages $V_{bd}$ of this structure for $N_{ox}$ values of $1\times 10^{12}$, $2\times 10^{12}$ and $3\times 10^{12}$ cm$^{-2}$ and for different resistivities of the silicon are given in table~\ref{tab:breakdownvsdose}.
 Shown are the results of 2D simulations in Cartesian coordinates (x,y) for the straight section of the guard rings, and in polar coordinates (r,z) for the corners.
 Given, that already $\sim$~600~000 grid points are required for the 2D simulations, a 3D simulation of the guard-ring structure is not feasible.
 The results show that the minimum breakdown voltage is 830~V for a resistivity of 3~k$\Omega\cdot$cm and $N_{ox}$ of $2\times 10^{12}$~cm$^{-2}$.
 The breakdown voltage is above 900~V for resistivities of 5 and 8~k$\Omega\cdot$cm.
 Additional simulations with $N_{ox}$ of $5\times 10^{10}$~cm$^{-2}$ show that in the specified resistivity range of 3 to 8~k$\Omega\cdot$cm the depletion region does not touch the scribe line for voltages below 990~V\@.
 With increasing $N_{ox}$ the extension of the depletion region towards the scribe line decreases.

\begin{table}[htpb]
	\begin{center}
		\begin{tabular}{|c||r|r|r|r|r|r|}\hline
		     & \multicolumn{2}{c|}{3 k$\Omega\cdot$cm}
		     & \multicolumn{2}{c|}{5 k$\Omega\cdot$cm}
		     & \multicolumn{2}{c|}{8 k$\Omega\cdot$cm}  \\ \hline
		    $N_{ox}$ [cm$^{-2}$] & 2D (x,y) & 2D (r,z) & 2D (x,y) & 2D (r,z) &
		                         2D (x,y) & 2D (r,z) \\ \hline\hline
		    $1\times 10^{12}$ & $>1100$ V & 1060 V & $>1100$ V & $>1100$ V &
		                        $>1100$ V & $>1100$ V \\ \hline
		    $2\times 10^{12}$ &  1000 V &  830 V &  1080 V &   910 V &
		                          950 V &   950 V \\ \hline
		    $3\times 10^{12}$ &  1010 V &  840 V & $>1100$ V & 910 V &
		                         1000 V & 960 V \\ \hline
		\end{tabular}
           \caption{Results of the 2D simulations in Cartesian and polar coordinates of the breakdown voltage as function of the surface density of oxide charges and bulk resistivity for the optimized guard-ring design.}
		\label{tab:breakdownvsdose}
	\end{center}
\end{table}

\subsection{Pixel}
\begin{figure}[htpb]
  \centering
  \includegraphics[width=0.45\linewidth]{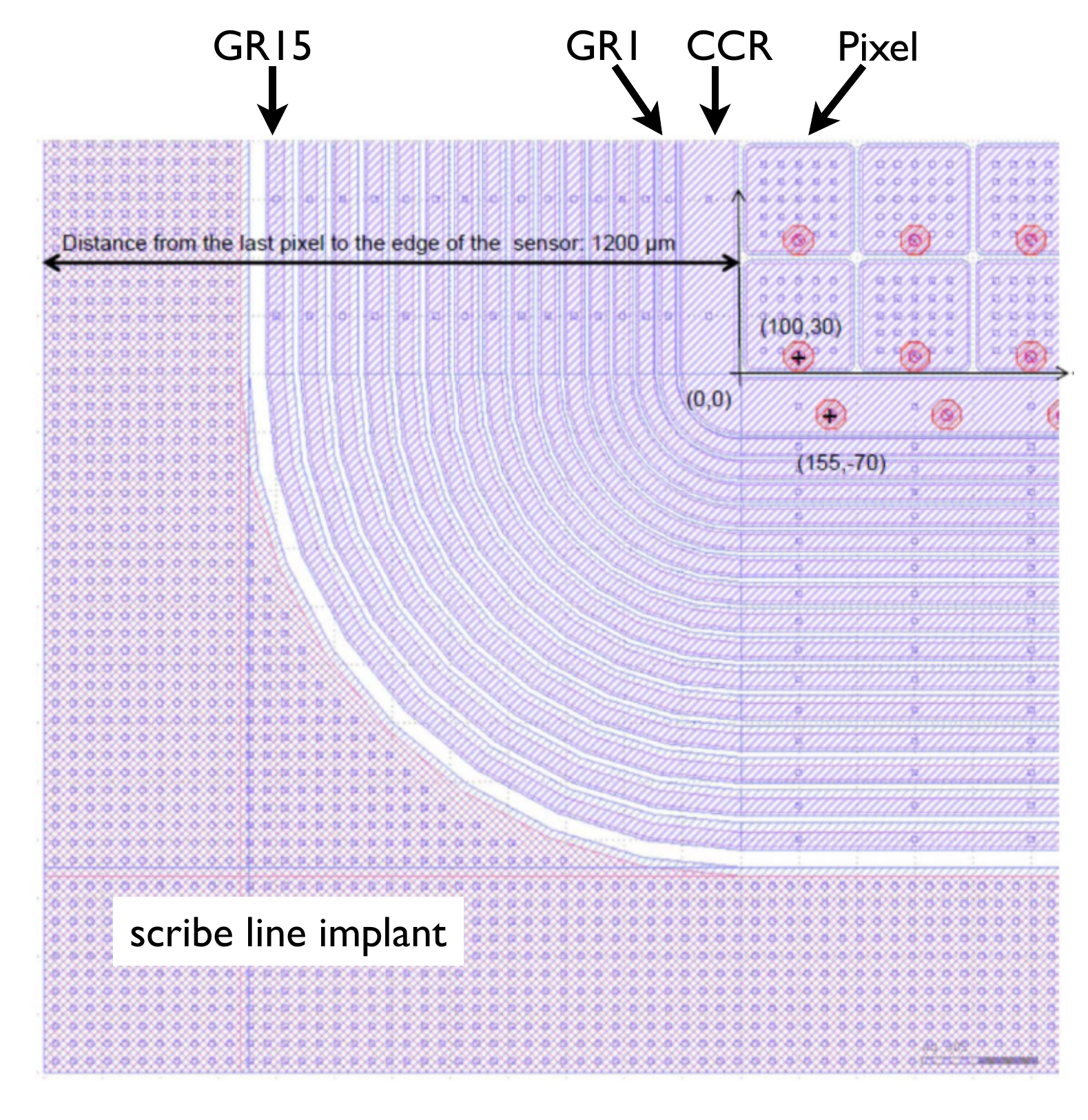}
  \hspace{0.5cm}
  \includegraphics[width=0.45\linewidth,height=13em]{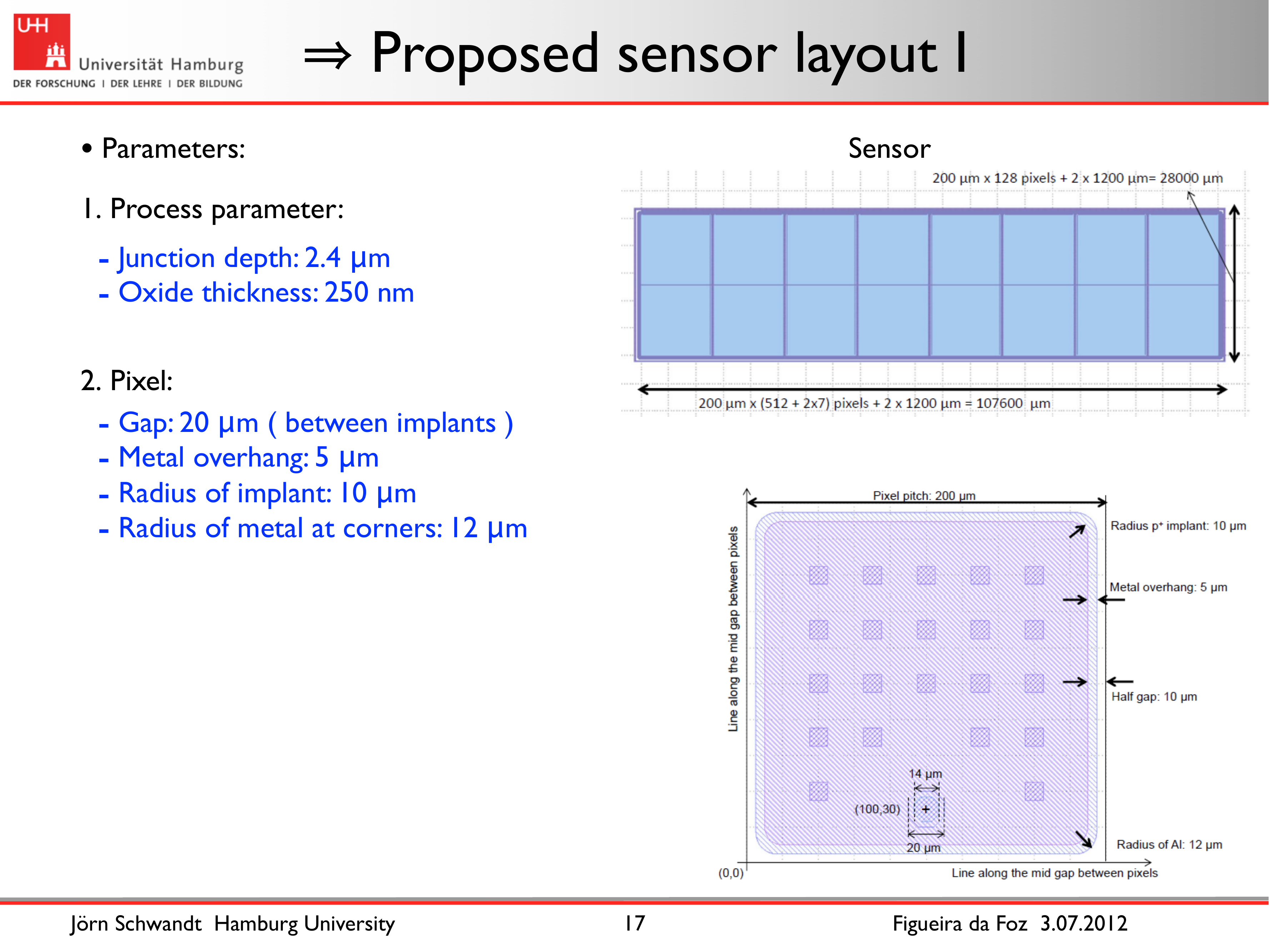}
  \caption{GDS printouts of the optimized layout of the AGIPD sensor.
  Left: From bottom left to top right one sees the scribe line, the scribe-line implant, the 15 guard rings (GR15 to GR1), the current-collection ring (CCR) and finally 6 pixels.
  Right: Pixel layout with the 21 vias which connect the Al to the $p^+$~implant, and the opening in the oxide for bump bonding (center bottom). At the periphery, the edge of the $p^+$-implantion widow and the Al overhang can be seen.}
  \label{fig:gds}
\end{figure}

 Because the pixel optimization is published in \cite{Schwandt:2012}, here we only show that with  above process parameters (junction depth of 2.4~$\upmu$m, oxide thickness 250~nm) and geometry parameters (gap of 20~$\upmu$m, Al overhang of 5~$\upmu$m, radius of junction of 10~$\upmu$m and radius of the Al layer of 12~$\upmu$m) the specification of table~\ref{tab:specs} are met.
 A printout of the GDS file of the optimized pixel layout is shown in figure~\ref{fig:gds} right.

\begin{figure}[htpb]
  \centering
  \includegraphics[width=0.7\linewidth]{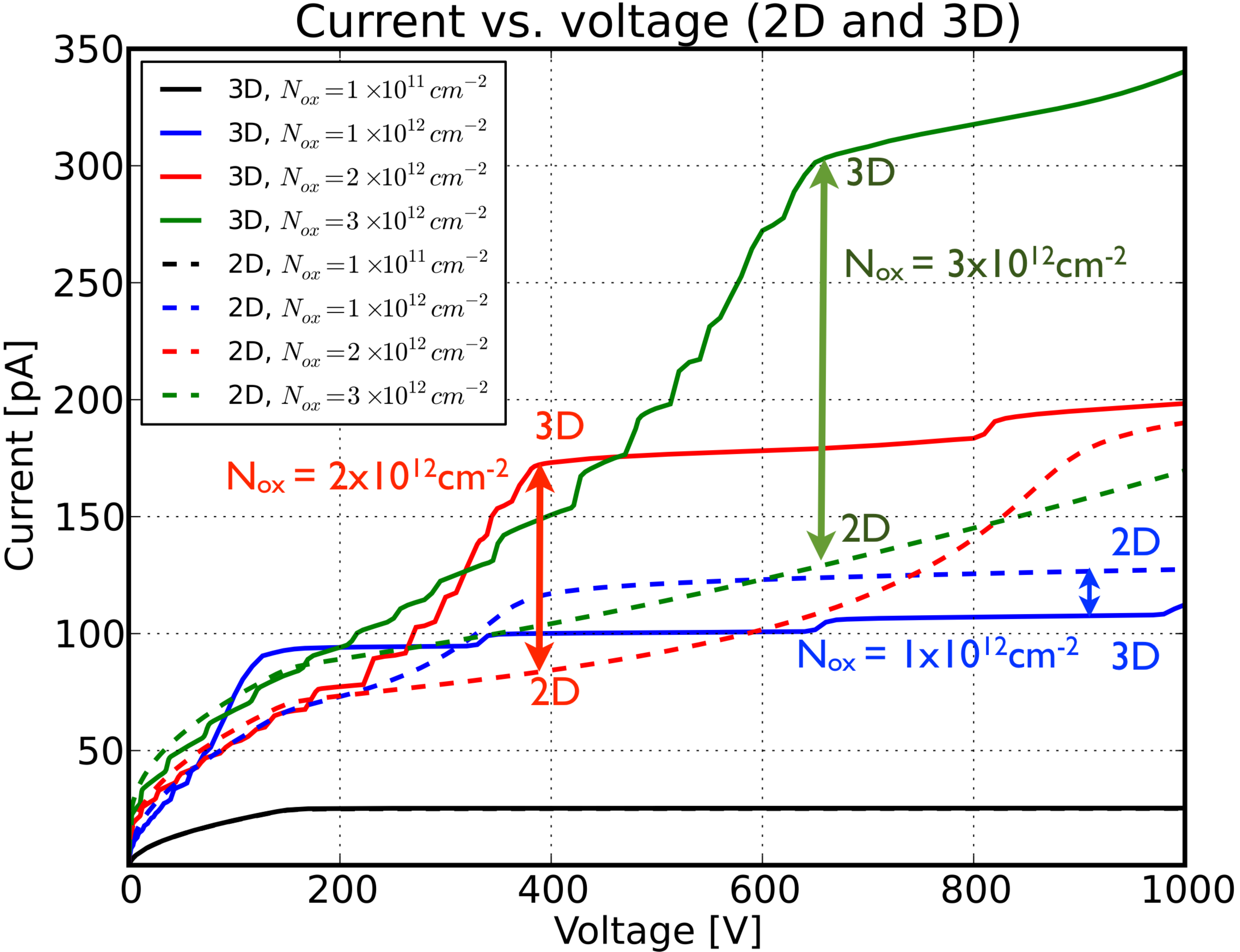}
   \caption{Comparison of the I-V curves for 2D simulations and for 3D simulations of a quarter of a pixel. The currents are scaled to a full pixel.}
  \label{fig:iv_2d_3d}
\end{figure}

 In figure~\ref{fig:iv_2d_3d} the I-V curves for 2D (x,y) simulations and for 3D simulations of a quarter of a pixel are plotted for the values of $N_{ox}$ and $J_{surf}$ given in table~\ref{tab:performance} for a resistivity of 5~k$\Omega\cdot$cm.
 The currents are scaled to the pixel size of 200 $\times$ 200 $\upmu$m$^2$.
 The breakdown voltage of the pixels for this design is above 1000~V.
 The difference in the voltage dependence of the currents between the 2D and the 3D simulations for high $N_{ox}$ values is due to the shape of the electron-accumulation layer at the corners of the pixel: The surface current is given by the product of the surface-current density times the area of the depleted surface at the Si-SiO$_2$ interface.
 The corners of the pixels can only be properly simulated in 3D\@.
The inter-pixel capacitance $C_{int}$, the sensor dark current $I_{sensor}$
and the CCR dark current $I_{CCR}$ are listed in table~\ref{tab:performance}
for a voltage of 500~V and different $N_{ox}$ values. All values are within the
specifications.
\begin{table}[htpb]
	\begin{center}
		\begin{tabular}{|c|c||c|c|c|} \hline
		    $N_{ox}$ & $J_{surf}$ & $I_{sensor}$ & $I_{CCR}$  & $C_{int}$ \\
$[cm^{-2}]$ &  $[\upmu$A/cm$^2]$ & $[\upmu$A$]$ & $[\upmu$A$]$ & [fF] \\ \hline\hline
       $1\times 10^{12}$ & 2 & 7.4  & 0.6 & 120 \\ \hline
       $2\times 10^{12}$ & 5 & 12.7 & 0.9 & 270 \\ \hline
       $3\times 10^{12}$ & 8 & 14.4 & 1.2 & 312 \\ \hline
		\end{tabular}\\
		\caption{Simulated values of the inter-pixel capacitance $C_{int}$, the sensor dark current $I_{sensor}$ and the CCR dark current $I_{CCR}$ at 500~V\@.}
		\label{tab:performance}
	\end{center}
\end{table}

\section{Summary}
 Experimental results on the dose dependence of the surface density of oxide charges and the surface-current density have been implemented in the SYNOPSYS TCAD simulation program and the design of the pixel and guard-ring layout for the AGIPD sensor has been optimized.
 The TCAD simulations show that the optimized sensor design meets the specifications for the AGIPD, in particular breakdown voltage, dark current and inter-pixel capacitance.

 \acknowledgments{This work was performed within the AGIPD Project financed by the European XFEL-Company. Additional support was provided by the Helmholtz Alliance \emph{Physics at the Terascale} and J. Zhang was supported by the Marie Curie Initial Training Network \emph{MC-PAD}. We thank the AGIPD colleagues for the excellent collaboration and the funding institutions for their support.
}

\end{document}